\documentclass[11pt]{article}
\usepackage{times,fullpage}  
\usepackage{mathrsfs,graphicx,amsmath,amssymb}
\usepackage[ruled,vlined]{algorithm2e}

\def\X{{\mathscr{X}}}
\def\F{{\mathscr{F}}}
\def\AC{{\mathscr{AC}}}

\newtheorem{observation}{Observation}

\newtheorem{definition}{Definition}

\newcommand{\QED}{\nobreak\hfill\hbox{$\square$}}

\title{Fast Algorithms for Reconciliation under Hybridization and Incomplete Lineage Sorting\thanks{This work was supported in part by NSF grant DBI-1062463, grant R01LM009494 from the National Library of Medicine, an Alfred P. Sloan Research Fellowship, and a Guggenheim Fellowship to L.N. The contents are solely the responsibility of the authors and do not necessarily represent the official views of the NSF, National Library of Medicine, the National Institutes of Health, the Alfred P. Sloan Foundation, or the John Simon Guggenheim Memorial Foundation. The authors wish to thank R. Matthew Barnett for providing the code for generating random species networks.
}}
\author
{Yun Yu and Luay Nakhleh\\
\\
\normalsize{Department of Computer Science, Rice University,}\\
\normalsize{6100 Main Street, Houston, TX 77005, USA}\\
\normalsize{E-mail:  \{yy9, nakhleh\}@cs.rice.edu.}
}

\date{}

\begin{document}

\maketitle 

\thispagestyle{empty}
\begin{abstract}
Reconciling a gene tree with a species tree is an important  task that reveals much about the evolution of genes, 
 genomes, and species, as well as about the molecular function of genes. A wide array of computational tools have been 
 devised for this task under certain evolutionary events such as hybridization, gene duplication/loss, or incomplete lineage 
 sorting. Work on reconciling gene tree with species phylogenies under two or more of these events have also begun to 
 emerge. Our group recently devised both parsimony and probabilistic frameworks for reconciling a gene tree with a phylogenetic 
 {\em network}, thus allowing for the detection of hybridization in the presence of incomplete lineage sorting. While the frameworks 
 were general and could handle any topology, they are computationally intensive, rendering their application to large datasets 
 infeasible. In this paper, we  present two novel approaches  to address the computational challenges of the two frameworks
 that are based on the concept of {\em ancestral configurations}. Our approaches still compute exact solutions while improving the 
 computational time by up to five orders of magnitude. These substantial gains in speed scale the applicability of these unified 
 reconciliation frameworks to much larger data sets. We discuss how the topological features of the gene tree and phylogenetic 
 network may affect the performance of the new algorithms. We have implemented the algorithms in our PhyloNet software 
 package, which is publicly available in open source. 
 \end{abstract}

\newpage

\setcounter{page}{1}

\section{Introduction}
\vspace{-.15in}
 Analysis of the increasingly available genomic data continue to reveal the extent of hybridization and its importance in the speciation  
 and evolutionary innovations of several groups of species and animals \cite{Arnold97, Barton01, Mallet05, Mallet07, Rieseberg97}. 
 When hybridization occurs, the evolutionary history of the species and their genomes is {\em reticulate} and best modeled by a 
 {\em phylogenetic network} which, in our context, is a special type of rooted, directed, acyclic graphs \cite{Nakhleh10}. Methods have 
 been devised for inferring phylogenetic networks from pairs of gene trees (e.g., \cite{EEEP, HorizStory, Nakhleh10, NakhlehRuths05}), 
 larger collections of gene trees \cite{Iersel:2009p22113,Wu:2010p22034,ParkNakhleh10,ParkNakhleh12}, and directly from 
 sequence data (e.g., \cite{GusfieldEtAl04,NakhlehJin05,JinNakhleh-bioinfo,JinNakhleh-mbe07,ParkMPPhylo,ParkNakhleh12a}).
 A salient feature of all these methods is that the incongruence of gene tree topologies, and more generally the heterogeneity among 
 the different loci, is caused solely by reticulate evolutionary events such as horizontal gene transfer or hybridization. 
 
 While hybridization causes incongruence among gene trees, other evolutionary events can also result in incongruence, such 
 as incomplete lineage sorting (ILS) and gene duplication/loss \cite{Maddison97}. In particular, as (successful) hybridization occurs 
 between closely related species, it is important to account simultaneously for incomplete lineage sorting, a phenomenon that 
 arises in similar situations \cite{LinderRieseberg04,Mallet05}. While a wide array of methods have been devised for inference under 
 ILS along (see \cite{DegnanRosenberg09,LiuEtAl09} for recent surveys), it is important to integrate both hybridization and ILS 
 into a single framework for inference. Needless to say, it is important to integrate all sources of incongruence into a single framework,
 but that is much beyond the scope of this paper. 
  The main task, then, becomes: given a gene tree topology and 
 a phylogenetic network, to reconcile the gene tree within the branches of the phylogenetic network, thus allowing simultaneously 
 for hybridization and ILS. When a method for achieving this task is ``wrapped" by a strategy for searching the phylogenetic network
 space, the result is a method for inferring reticulate evolutionary histories in the presence of both hybridization and ILS. Therefore, 
 is is very important to solve the reconciliation problem. 
 
 Indeed, in the last five years, several attempts have been made, following different approaches, to address the problem of inferring 
  hybridization in the presence of ILS \cite{ThanEtAl07, HollandEtAl08, MengKubatko09, Kubatko09, JolyEtAl09, YuEtAl11a}. 
  However, due to the computational challenges of the problem, these methods focused on very limited cases: fewer than 5 taxa, one 
  or two hybridization events, and a single allele sampled per species. More recently, our group proposed two methods for detecting 
  hybridization in the presence of incomplete lineage sorting, including a probabilistic method which computes the probability of gene 
  tree topologies given a phylogenetic network \cite{YuEtAl12a} and a parsimony method which computes the minimum number of 
  extra lineages \cite{Maddison97} required to reconcile a gene tree within the branches of a phylogenetic network \cite{YuEtAl12b}. 
  While these methods are general in terms of the topologies and sizes of gene trees and phylogenetic networks, they are 
  computationally intensive. In particular, these methods convert a phylogenetic network to a special type of trees, called 
  {\em multil-labeled trees} (MUL-trees), and conduct computation on these trees while accounting for every possible mapping of 
  genes to their leaves. This computation can be exponential in the number of leaves, and does explicit computations of 
  coalescent histories of the gene genealogies. 
  
In this paper, we propose a novel way of computing the probability of gene tree topologies given a phylogenetic network, and a novel way of computing the minimum number of extra lineages of a gene tree and a phylogenetic network. Both of them use the concept of \textit{ancestral configuration} (or AC) which was introduced very recently for computing the probability of gene tree topologies given a species tree \cite{Wu12}. The new algorithms are exact and  much more efficient than the two MUL-tree based algorithms we introduced in \cite{YuEtAl12a,YuEtAl12b}.  In our extensive simulation studies, we compared the running time of the new AC-based methods with the previous MUL-tree based ones. We show that the new algorithms can speed up the computation by up to 5 orders of magnitude, thus allowing for the analysis of much larger data sets. Furthermore, we discuss how the running time of the new methods  is still affected by the topologies of the species networks, more specifically the configurations of reticulation nodes, and the topologies of gene trees. All methods described in this paper have been implemented in the PhyloNet software package \cite{ThanEtAl08} which is freely available for download in open source at http://bioinfo.cs.rice.edu/phylonet.
 
 \vspace{-.2in}
\section{Background}
 \vspace{-.2in}
In this work, we assume the following definition of phylogenetic networks \cite{Nakhleh10}.
 \vspace{-.1in}
\begin{definition}
 \label{net-def}
 A {\em phylogenetic $\X$-network}, or $\X$-network for short, $N$ is an ordered pair $(G,\ell)$, where
 $G=(V,E)$ is a directed, acyclic graph (DAG) with $V=\{r\} \cup V_L \cup V_T \cup V_N $, where
       (1) $indeg(r)=0$ ($r$ is the {\em root} of $N$);
        (2) $\forall{v \in V_L}$, $indeg(v)=1$ and $outdeg(v)=0$ ($V_L$ are the {\em external tree nodes}, or {\em leaves}, of $N$);
        (3) $\forall{v \in V_T}$, $indeg(v)=1$ and $outdeg(v) \geq 2$ ($V_T$ are the {\em internal tree nodes} of $N$); and, 
        (4) $\forall{v \in V_N}$, $indeg(v) = 2$ and $outdeg(v) = 1$ ($V_N$ are the {\em reticulation nodes}
         of $N$);
        $E \subseteq V \times V$ are the network's edges , and 
  $\ell:V_L \rightarrow \X$ is the {\em leaf-labeling} function, which is a bijection from $V_L$ to 
 $\X$. 
\end{definition}
 For the probabilistic setting of the problem, we also associate with every pair of reticulation edges inheritance probabilities $\gamma_{(u_1,v)}$ and $\gamma_{(u_2,v)}$ such that $\gamma_{(u_1,v)}+\gamma_{(u_2,v)}=1$. Inheritance probability $\gamma_{(u,v)}$ indicates the proportion of alleles in population $v$ that are inherited from population $u$. A {\em gene tree} is 
 a phylogenetic network with no reticulation nodes. 
 
 The way in which a gene evolves within the the branches of a phylogenetic network can be described by a {\em coalescent 
 history} \cite{YuEtAl12a}.  Let $N$ be a phylogenetic network. We denote by $V(N)$ the set of nodes in $N$ and
  by $N_u$ the set of nodes that are reachable from the root of $N$ via at least one path that goes through node $u \in V(N)$. Given 
  a phylogenetic network $N$ and a gene tree $g$, a  \textit{coalescent history} is a function $h:V(g) \rightarrow V(N)$ such that the following two conditions hold:
 (1) if $v$ is a leaf in $g$, then $h(v)$ is the leaf in $N$ with the same label (in the case of multiple alleles, $h(v)$ is the leaf in $N$ with the label of the species from which the allele labeling leaf $v$ in $g$ is sampled); and, 
(2) if $v$ is a node in $g_u$, then $h(v)$ is a node in $N_{h(u)}$. 
 See Fig.~\ref{fig:CoalHis} in the Appendix for an illustration. 

Given a phylogenetic network $N$ and a gene tree $g$, we denote by $H_N(g)$ the set of all coalescent histories. Then the probability of observing gene tree $g$ given phylogenetic network $N$ is 
\begin{equation}
\label{eq:probUsingCH}
P(g|N)=\sum_{h \in H_N(g)}P(h|N),
\end{equation}
where $P(h|N)$ is the probability of coalescent history $h$ given phylogenetic network $N$ (along with its branch lengths 
and inheritance probabilities). 
%
 Coalescent histories can also be used to compute the minimum number of extra lineages required to reconcile gene tree $g$ with  
  $N$, which we denote by $XL(N,g)$, as 
\begin{equation}
\label{eq:xlUsingCH}
XL(N,g)=\min_{h \in H_N(g)}XL(N,h),
\end{equation}

Methods for computing $P(g|N)$ and $XL(N,g)$ when $N$ is a tree were recently given in \cite{DegnanSalter05} and \cite{ThanNakhleh-PLoSCB09},
respectively. Recently, we proposed new methods for computing these two quantities when $N$ is a phylogenetic network
\cite{YuEtAl12a,YuEtAl12b}. The basic idea of both of these methods is to convert the phylogenetic network $N$ into a \textit{MUL-tree} $T$ and then make use of some existing techniques to complete the computation on $T$ instead of on $N$. A MUL-tree \cite{HuberOxelman06} is a tree whose leaves are not uniquely labeled by a set of taxa. Therefore, alleles sampled from one 
 species, say $x$, can map to any of the leaves in the MUL-tree $T$ that are labeled by $x$. For network $N$ on taxa $\X$, we denote by $A_x$ the set of alleles sampled from species $x$ ($x \in \X$), and by $c_x$ the set of leaves in $T$ that are labeled by species $x$. Then a \textit{valid allele mapping} is a function $f:(\cup_{x \in \X}A_x) \rightarrow (\cup_{x \in \X}c_x)$ such that if $f(a) = d$, and $d \in c_x$, then $a \in A_x$ \cite{YuEtAl12a}. Fig. \ref{fig:Net2Tree} in the Appendix shows an example of converting a phylogenetic network into a MUL-tree along with all valid allele mappings when single allele is sampled per species.

Suppose $T$ is the MUL-tree converted from network $N$. We denote by $\F_{T,g}$ the set of all valid allele mappings for MUL-tree $T$ and gene tree $g$. Then the probability of observing gene tree $g$ given $N$ can be computed using MUL-tree $T$ as follows
\begin{equation}
\label{eq:probUsingMul}
P(g|N)=\sum_{f \in \F_{T,g}}{\sum_{h \in H_{T,f}(g)}P(h|T,f)},
\end{equation}
where $H_{T,f}(g)$ is the set of coalescent histories of $g$ within MUL-tree $T$ under valid allele mapping $f$, and $P(h|T,f)$ is the probability of observing coalescent history $h$ within $T$ under $f$ \cite{YuEtAl12a}. Furthermore, the minimum number of extra lineages required to reconcile gene tree $g$ with $N$ can also be computed using MUL-tree $T$ by
\begin{equation}
\label{eq:xlUsingMul}
XL(N,g)=\min_{f \in \F_{T,g}}{\min_{h \in H_{T,f}(g)}XL(T,f,h)},
\end{equation}
where $XL(T,f,h)$ is the total number of extra lineages of coalescent history $h$ within $T$ under allele mapping $f$  \cite{YuEtAl12b}.

The advantage of the MUL-tree based techniques is that once the network is converted to the MUL-tree, tree-based techniques from
the multi-species coalescent theory apply with minimal revision. Nonetheless, from Eq. \eqref{eq:probUsingMul} and Eq. \eqref{eq:xlUsingMul} we can see that the running time of both two methods depend on the number of valid allele mappings. Let $V_L(N)$ be the set of leaves of $N$, and $a(x)$ be the number of alleles sampled from some $x$ in $V_L(N)$ in gene tree $g$. Then the number of valid allele mappings of $N$ and $g$ is bounded from below and above by $2^{\sum_{x \in V_L(N)}{r_{min}(x)a(x)}}$ and $2^{\sum_{x \in V_L(N)}{r_{max}(x)a(x)}}$, respectively, where  $r_{min}(x)$ and $r_{max}(x)$  are the minimum and maximum number of reticulation nodes on any path from leaf $x$ in $V_L(N)$ to the root of $N$ respectively. We can see that when the number of taxa or sampled alleles increases, or when the number of reticulation nodes increases, this number can quickly become very large which makes the computations prohibitive. Furthermore, computing term $P(h|T,f)$ in Eq. \eqref{eq:probUsingMul} using coalescent histories will become infeasible when the number of taxa or sampled alleles increases \cite{Wu12}. 

\vspace{-.2in}
\section{Methods}
\begin{figure*}[ht]   
  \begin{center}
   \vspace{-2mm}
   \includegraphics[width=0.9\textwidth]{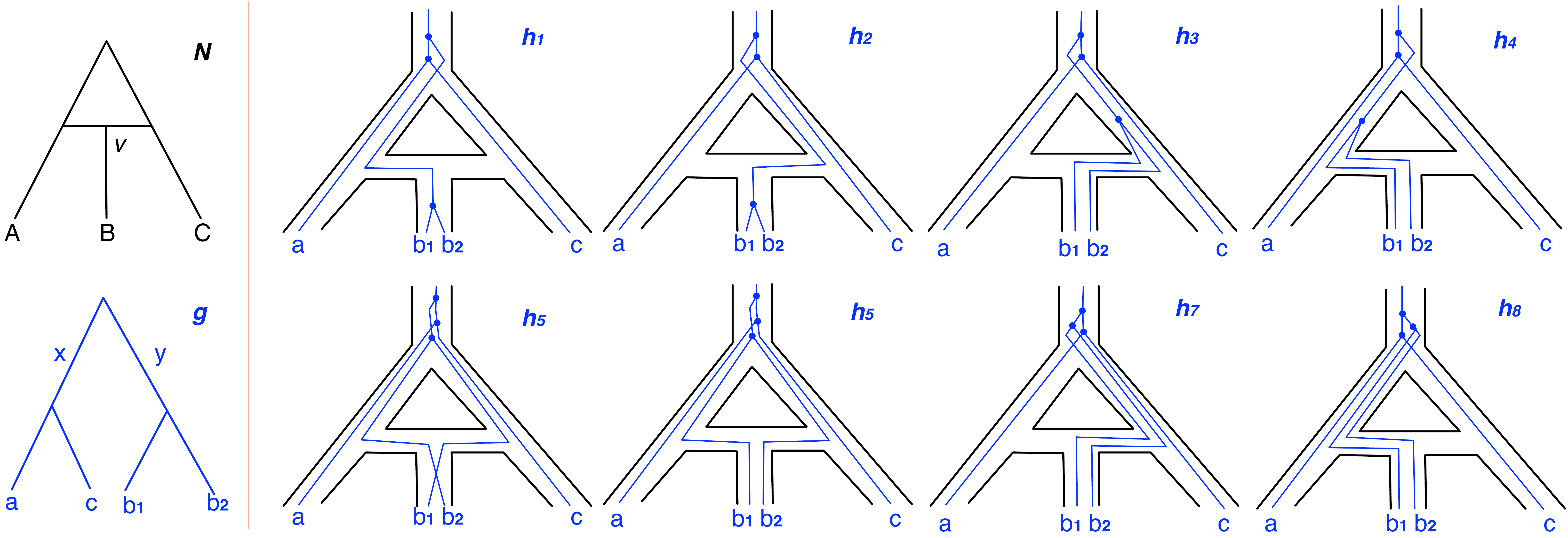}
  \renewcommand{\baselinestretch}{1.0}
   \vspace{-3mm}
   \caption{\small A phylogenetic network $N$, a gene tree $g$, and the eight possible coalescent histories of $g$ within the 
   branches of $N$. Here, one allele is sampled from taxa A and C, and two alleles from taxon B. \label{fig:CoalHis} 
   }
\vspace{-4mm}
  \end{center}
\end{figure*}

 \vspace{-.15in}
Central to our methods is the concept of \textit{ancestral configuration} (or simply configuration, or AC). When it was first introduced, it was defined on species trees for computing the probability of gene tree topologies \cite{Wu12}. In this work, we extend it to species networks. Given a species network $N$ and a gene tree $g$, an ancestral configuration at node $v$ of $N$, which we denote by $AC_v$ (the subscript $v$ may be omitted when the identity of node $v$ is clear from the context), is a set of gene lineages at node $v$ under some coalescent history $h$ in $H_N(g)$. The number of gene lineages in configuration $AC_v$ is denoted by $n(AC_v)$. For example, given the coalescent history $h_3$ shown in Fig. \ref{fig:CoalHis}, for reticulation node $v$, we have $AC=\{b_1,b_2\}$ and $n(AC)=2$; for the root of $N$, we have $AC=\{a,c,y\}$ and $n(AC)=3$. Furthermore, we denote by $\AC_v$ a set of pairs $(a,w)$ where $a$ is a configuration at node $v$ of $N$ and $w$ is the weight of $a$, and by $\AC_{(u,v)}$ a set of $(a,w)$ where $a$ is a configuration that about to leave branch $(u,v)$ of $N$ and $w$ is the weight of $a$. We will discuss how to set/use the weight $w$ below. 

Assume $m$ and $n$ are two gene lineages that meet at some node in a gene tree $g$. When reconciling $g$ within the branches of a species network $N$, after they two entered the same branch of $N$, they might or might not have coalesced before leaving that branch, the probability of which depends on the length (in terms of time) and width (in terms of population size) of that branch. Therefore, one configuration entering a branch of $N$ might give rise to several different configurations leaving that branch with different probabilities. For example, suppose a gene tree $g$ has a subtree $((a,b)x,c)y$ (tree with root $y$, leaf-child $c$ of the 
root, child $x$ of the root, and two leaves $a$ and $b$ that are children of $x$). Then if a configuration $\{a,b,c\}$ entered a branch of $N$, it could give rise to one of three different configurations leaving that branch, including $\{a,b,c\}$ $\{x,c\}$ and $\{y\}$. We 
 denote by $Coal(AC,g)$, for configuration $AC$ and gene tree $g$, the set of all configurations that $AC$ might coalesce into with respect to the topology of $g$. 
We now show  how to use configurations to compute  $P(g|N)$ and $XL(N,g)$ efficiently. 
 
 \vspace{-.2in}
\subsection{Counting the number of extra lineages}
\vspace{-.1in}
For a configuration $AC$, we denote by $xl(AC)$ the minimum total number of extra lineages on all branches that the extant gene lineages in $AC$ having passed through from time $0$ to coalesce into the present gene lineages in $AC$. In this method, weight $w$ in $(AC,w) \in \AC$ corresponds to $xl(AC)$, where $\AC$ is either $\AC_v$ where $v$ is a node or $\AC_b$ where $b$ is a branch.
\begin{observation}
\label{lemma:xlUpdate}
Let $AC$ be a configuration entering a branch $b$ and $AC^+$ be a configuration that $AC$ coalesced into when leaving $b$. Then 
\begin{equation}
\label{eq:xlUpdate}
xl(AC^+)=xl(AC)+n(AC^+)-1,
\end{equation}
where $n(AC^+)-1$ is the number of extra lineages on branch $b$.
\end{observation}

We define a function called {\bf CreateCACsForXL} which takes a gene tree $g$, a branch $b=(u,v)$ of the network $N$ and a set of configuration-weight pairs $\AC_v$ that enter branch $b$, and returns a set of configuration-weight pairs $\AC_{(u,v)}$ that leave branch $b$. 

 \begin{algorithm}[ht]
\footnotesize
  \KwIn{Gene tree $g$, a branch $b=(u,v)$, a set of configuration-weight pairs $\AC_v$}
 \KwOut{A set of configuration-weight pairs $\AC_{(u,v)}$}
 \ForEach{$(AC,xl(AC)) \in \AC_v$}{
	$AC^+ \leftarrow \text{argmin}_{AC' \in Coal(AC,g)}{n(AC')}$\; 
	Compute $xl(AC^+)$ using Eq. \eqref{eq:xlUpdate}\;
	$\AC_{(u,v)} \leftarrow \AC_{(u,v)} \cup (AC^+, xl(AC^+))$ \;
 }
 \Return{$\AC_{(u,v)}$}
  \caption{{\bf CreateCACsForXL.} \label{alg:CreateCACsForXL}}
 \end{algorithm}
Note that although one configuration can coalesce into several different configurations along a branch, under parsimony we only need to keep the one that has the minimum total number of extra lineages. Therefore $|\AC_v|=|\AC_{(u,v)}|$ and there is 1-1 correspondence between configurations in $|\AC_v|$ and configurations in $|\AC_{(u,v)}|$.

For a phylogenetic network $N$ and a gene tree $g$, the algorithm for computing the minimum number of extra lineages required to reconcile $g$ within $N$ is shown in Alg. \ref{alg:countXL}. Basically, we traverse the nodes of the network in post-order. For every node $v$ we visit, we construct the set of configuration-weight pairs $\AC_v$ for node $v$ based on its type. Recall that there are four types of nodes in a phylogenetic network, which are leaves, reticulation nodes, internal tree nodes, and the root. Finally when we arrive at the root of $N$, we are able to obtain $XL(N,g)$.

 \begin{algorithm}[ht]
 \footnotesize
  \KwIn{Phylogenetic network $N$, gene tree $g$}
 \KwOut{$XL(N,g)$}
 \While{traversing the nodes of $N$ in post-order}{ 
 \If{node $v$ is a leaf, who has parent $u$}{
 	$\AC_v \leftarrow \{(AC,0)\}$ where $AC$ is the set of leaves in $g$ sampled from the species $v$ which is labeled by\;
   	$\AC_{(u,v)} \leftarrow$ CreateCACsForXL($g,(u,v),\AC_v$)\;
 }
 \ElseIf{node $v$ is a reticulation node, who has child $w$, and two parents $u_1$ and $u_2$}{
   	$\AC_v \leftarrow \AC_{(v,w)}$\;
   	\ForEach{$(AC,xl(AC)) \in \AC_v$}{
   		\ForEach{$AC_1 \subseteq AC$}{
	 		$AC_2 \leftarrow AC-AC_1$\;
			$\AC_{(u_1,v)} \leftarrow \AC_{(u_1,v)} \cup (AC_1, xl(AC)+n(AC_1)-1)$\;
			$\AC_{(u_2,v)} \leftarrow \AC_{(u_2,v)} \cup (AC_2, n(AC_2)-1)$\;
		}			
  	 }
 }
  \ElseIf{node $v$ is an internal tree node or root, who has two children $w_1$ and $w_2$}{
  	\ForEach{$(AC_1,xl(AC_1)) \in \AC_{(v,w_1)}$}{
		\ForEach{$(AC_2,xl(AC_2)) \in \AC_{(v,w_2)}$}{
			\If{$AC_1$ and $AC_2$ are \textit{compatible}}{
				$\AC_v \leftarrow \AC_v \cup (AC_1 \cup AC_2, xl(AC_1)+xl(AC_2))$
			}
		}
	}
	\If{node $v$ is an internal tree node, who has a parent $u$}{
		$\AC_{(u,v)} \leftarrow$ CreateCACsForXL($g,(u,v),\AC_v$)\;	
	}
	\Else{
		\Return {$\min_{(AC, xl(AC)) \in \AC_v}xl(AC)$}\;
		
	}	
 }
 }
 \caption{{\bf CountXL.} \label{alg:countXL}}
 \end{algorithm}
 
At a reticulate node $v$ who has parents $u_1$ and $u_2$, every gene lineage could independently choose to go toward $u_1$ or $u_2$. So for every $(AC,xl(AC))$ in $\AC_v$, there are $2^{n(AC)}$ different ways of splitting $AC$ into two configurations, say $AC_1$ and $AC_2$, such that $AC_v=AC_1 \cup AC_2$. For example, a configuration $\{a,b\}$ can be split in four different ways including $\{a,b\}$ and $\{\emptyset\}$, $\{a\}$ and $\{b\}$, $\{b\}$ and $\{a\}$, and $\{\emptyset\}$ and $\{a,b\}$. It is important to keep track of those gene lineages that are originally coming from one splitting so that we could merge them back once they are in the same population again. Note that there is no need to consider coalescent events on branch $(u_1,v)$ or $(u_2,v)$, because all gene lineages on these two branches were already in the same population on branch $(v,w)$. And under parsimony where all gene lineages are assumed to coalesce as soon as they can, all possible coalescent events that could happen among these gene lineages must have already been applied on branch $(u,w)$. As a result, $(AC_1,xl(AC_1))$ and $(AC_2,xl(AC_2))$ can be put directly into $\AC_{(u_1,v)}$ and $\AC_{(u_2,v)}$ respectively. 

The {\em compatibility} of configurations in the algorithm is defined as follows. Two configurations are compatible if  for 
every reticulation node, either both configurations went through that node and had resulted from the same split of an 
ancestral configuration, or at least one of the two configurations did not go through that node. 
 Fig. \ref{fig:algoExample} in the Appendix illustrates configurations generated for every node and branch of a network given a gene tree. 
%

\vspace{-.2in}
\subsection{Calculating gene tree probability}
\vspace{-.1in}
For a configuration $AC$, we denote by $p(AC)$ the cumulative probability of the extant gene lineages in $AC$ coalescing into the present gene lineages in $AC$ from time $0$. In this method, weight $w$ in $(AC,w) \in \AC$ corresponds to $p(AC)$, where $\AC$ is either $\AC_v$ where $v$ is a node or $\AC_b$ where $b$ is a branch.
\begin{observation}
\label{lemma:coalesceProbOnBranch}
Let $AC$ be a configuration entering branch $b$ of network $N$ with branch length $\lambda_b$. Then the probability of observing configuration $AC^+$ leaving branch $b$ is
\begin{equation}
\label{eq:coalesceProbOnBranch}
p_t(AC, AC^+, b) = p_{n(AC),n(AC^+)}(\lambda_b)\frac{w_b(AC,AC^+)}{d_b(AC,AC^+)},
\end{equation}
where $p_{n(AC),n(AC^+)}(\lambda_b)$ is the probability that $n(AC)$ gene lineages coalesce into $n(AC^+)$ gene lineages within time $\lambda_b$, $w_b(AC,AC^+)$ is the number of ways that coalescent events can occur along branch $b$ to coalesce $AC$ into $AC^+$ with respect to the gene tree topology, and $d_b(AC,AC^+)$ is the number of all possible orderings of $n(AC)-n(AC^+)$ coalescent events. 
\end{observation}
The details of how to compute  $p_{n(AC),n(AC^+)}(\lambda_b)$, $w_b(AC,AC^+)$ and $d_b(AC,AC^+)$ are given in \cite{DegnanSalter05}.
\begin{observation}
\label{lemma:probUpdate}
Let $AC$ be a configuration entering a branch $b$ and $AC^+$ be a configuration that $AC$ coalesced into when leaving $b$. Then 
\begin{equation}
\label{eq:probUpdate}
p(AC^+)=p(AC)p_t(AC,AC^+,b).
\end{equation}
\end{observation}

We define a function called {\bf CreateCACsForProb} which takes a gene tree $g$, a branch $b=(u,v)$ of the network $N$ and a set of configuration-weight pairs $\AC_v$ that enter branch $b$, and returns a set of all possible configuration-weight pairs $\AC_{(u,v)}$ that leave branch $b$.

 \begin{algorithm}[ht]
 \footnotesize
  \KwIn{Gene tree $g$, a branch $b=(u,v)$, a set of configuration-weight pairs $\AC_v$}
 \KwOut{A set of configuration-weight pairs $\AC_{(u,v)}$}
 \ForEach{$(AC,p(AC)) \in \AC_v$}{
 	$S \leftarrow $ Coal$(AC,g)$\;
	\ForEach{$AC^+ \in S$}{
		Compute $p(AC^+)$ using Eq. \eqref{eq:probUpdate}\;
		\If{$(AC^+, w) \in \AC_{(u,v)}$ for some weight $w$}{
			$w \leftarrow w+p(AC^+)$			
		}
		\Else{
			$\AC_{(u,v)} \leftarrow \AC_{(u,v)} \cup (AC^+, p(AC^+))$ \;
		}
	} 
 }
 \Return{$\AC_{(u,v)}$}\;
  \caption{{\bf CreateCACsForProb.} \label{alg:CreateCACsForML}}
 \end{algorithm}

\begin{figure*}[ht]   
  \begin{center}
   \vspace{-2mm}
   \includegraphics[width=0.8\textwidth]{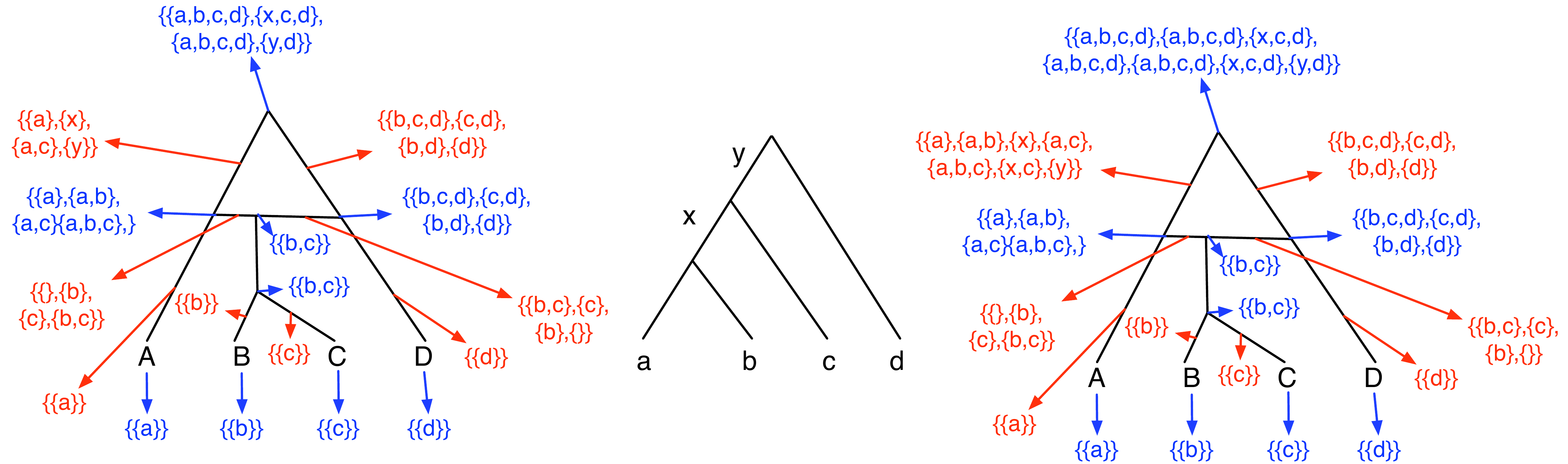}
  \renewcommand{\baselinestretch}{1.0}
   \vspace{-3mm}
   \caption{\small (Left) A phylogenetic  network with configurations generated when counting the minimum number of extra lineages of the gene tree in the middle. (Right) A phylogenetic network with configurations generated when computing the probability of the topology of the gene tree in the middle. Configurations (along with arrows) in blue represent configurations generated for nodes and configurations (along with arrows) in red represent configurations generated for branches. The weight of every configuration is not included in the figure. The root has two configurations $\{a,b,c,d\}$, but they are not the same because one represents the scenario where $a$ went left and $b$, $c$ and $d$ went right at the reticulation node, and the other represents the scenario where $a$ and $c$ went left and $b$ and $d$ went right at the reticulation node. \label{fig:algoExample} 
   }
\vspace{-4mm}
  \end{center}
\end{figure*}

Note that several configurations can coalesce into the same configuration along a branch, but we only need to keep one copy of every distinct configuration. Here, we define two configurations to be the \textit{identical} if they satisfy the following two conditions: (1) they contain the same set of gene lineages, and (2) for every reticulation node $v'$ in the network, either neither of them contain lineages that have passed through it, or the lineages in these two configurations that passed through it originally came from one splitting at node $v'$. 
 The algorithm for calculating the probability of observing a gene tree $g$ given a species network $N$ is shown in Alg. \ref{alg:calProb}. The basic idea is similar to the parsimony method we described in the previous section. An illustration is given in  Fig. \ref{fig:algoExample}.

 \begin{algorithm}[ht]
 \footnotesize
  \KwIn{Phylogenetic network $N$ including topology, branch lengths and inheritance probabilities, gene tree $g$}
 \KwOut{$P(g|N)$}
 \While{traversing the nodes of $N$ in post-order}{ 
 \If{node $v$ is a leaf, who has parent $u$}{
 	 $\AC_v \leftarrow \{(AC,1)\}$ where $AC$ is the set of leaves in $g$ sampled from the species which $v$ is labeled by\;
   	$\AC_{(u,v)} \leftarrow$ CreateCACsForProb($g,(u,v),\AC_v$)\;
 }
 \ElseIf{node $v$ is a reticulation node, who has child $w$, and two parents $u_1$ and $u_2$}{
   	$\AC_v \leftarrow \AC_{(v,w)}$\;
	$S_1 \leftarrow \emptyset$, $S_2 \leftarrow \emptyset$ \; 
   	\ForEach{$(AC,p(AC)) \in \AC_v$}{
   		\ForEach{$AC_1 \subseteq AC$}{
			$AC_2 \leftarrow AC-AC_1$\; 
			$S_1 \leftarrow (AC_1,p(AC) \gamma_{(u_1,v)}^{n(AC_1)})$\;
			$S_2 \leftarrow (AC_2,\gamma_{(u_2,v)}^{n(AC_2)})$\;
		}
  	 }
	 $\AC_{(u_1,v)} \leftarrow$ CreateCACsForProb($g,(u_1,v),S_1$)\;	
	 $\AC_{(u_2,v)} \leftarrow$ CreateCACsForProb($g,(u_2,v),S_2$)\;	
 }
  \ElseIf{node $v$ is an internal tree node or root, who has two children $w_1$ and $w_2$}{
  	\ForEach{$(AC_1,p(AC_1)) \in \AC_{(v,w_1)}$}{
		\ForEach{$(AC_2,p(AC_2)) \in \AC_{(v,w_2)}$}{
			\If{$AC_1$ and $AC_2$ are compatible}{
				$\AC_v \leftarrow \AC_v \cup (AC_1 \cup AC_2, p(AC_1)p(AC_2))$\;
			}
		}
	}
	\If{node $v$ is an internal tree node, who has a parent $u$}{
		$\AC_{(u,v)} \leftarrow$ CreateCACsForProb($g,(u,v),\AC_v$)\;	
	}
	\Else{
		Let $AC_R$ be the root lineage of the gene tree $g$ \;
		\Return {$\sum_{(AC,p(AC)) \in \AC_v}p_t(AC,AC_R,+\infty)p(AC)$}\;
		
	}	
 }
 }
 \caption{{\bf CalProb.} \label{alg:calProb}}
 \end{algorithm}

\vspace{-.2in}
\subsection{Reducing the number of configurations}
\vspace{-.1in}
At every reticulation node $v$ in the species network, every configuration $AC$ in $(AC,w) \in \AC_v$ is split into two configurations in all $2^{n(AC)}$ possible ways. This may result in multiple $(AC, w)$ pairs in a $\AC$ set where their configurations have the same set of gene lineages but not considered to be the same because they were not originally from one splitting at some reticulation node some lineages in them have passed through. It may increase the number of configurations significantly. It is clear that the running time of both these two algorithms depends on the number of configurations. So in order to reduce the number of configurations so as to speedup the computation, we make use of {\em articulation} nodes in the graph (an articulation node is a node whose removal 
disconnects the phylogenetic network). Obviously, the reticulation nodes inside the sub-network rooted at an articulation node are independent of the reticulation nodes outside the sub-network. So at articulation node $v$ we can clear all the information about the splittings at all reticulation nodes under $v$ so that all configurations at $v$ containing the same set of gene lineages are considered to be the same. More precisely, when traversing the species network, after constructing $\AC_v$ for some internal tree node $v$ as we have described in Alg. \ref{alg:countXL} and Alg. \ref{alg:calProb}, if $v$ is an articulation node, we clear all the information about splittings at all reticulation nodes in the sub-network rooted at $v$. 
Then for counting the minimum number of extra lineages, we update $\AC_v$ to be $\AC_v'$ such that only the configuration-weight pair that has the minimum weight is left, using the statement: 
$\AC_v' = \{\text{argmin}_{(AC,xl(AC) \in \AC_v}xl(AC)\}$.
 And for computing the probability of the topology of a gene tree, we keep only one copy of every distinct configuration in the sense of the set of lineages it contains. More precisely, we update $\AC_v$ to be $\AC_v'$ using
$\AC_v' = \{(AC, w'): w'=\sum_{(AC,w) \in \AC_v}w\}$.

\vspace{-.2in}
\section{Results and Discussion}
\vspace{-.15in}
To study the performance of the two methods compared to the MUL-tree based ones, we ran all four on synthetic data generated 
as follows. We first generated $100$ random 24-taxon species trees using PhyloGen \cite{Phylogen}, and from these we 
 generated  random species networks with $1$, $2$, $4$, $6$ and $8$ reticulation nodes. When expanding a species network with $n$ reticulation nodes to a species network with $n+1$ reticulation nodes, we randomly selected two existing edges in the species network and connected their midpoints from the higher one to the lower one and then the lower one becomes a new reticulation node. Then, we simulated $10$, $20$, $50$, $100$, $200$, $500$ and $1000$ gene trees respectively within the branches of each species network using the {\tt ms} program \cite{Hudson02}. 
  Since the MUL-tree methods are computationally very intensive, we employed the following strategy: for the parsimony methods,
 we bounded the time at 24 hours (that is, killed jobs that did not complete within 24 hours). For the probabilistic ones, we bounded 
 the time at 8 hours.  All computations were run on a computer with a quad-core Intel Xeon, 2.83GHz CPU, and 4GB of RAM.

\begin{figure*}[!ht]   
  \begin{center}
   \vspace{-2mm}
   \includegraphics[width=0.8\textwidth]{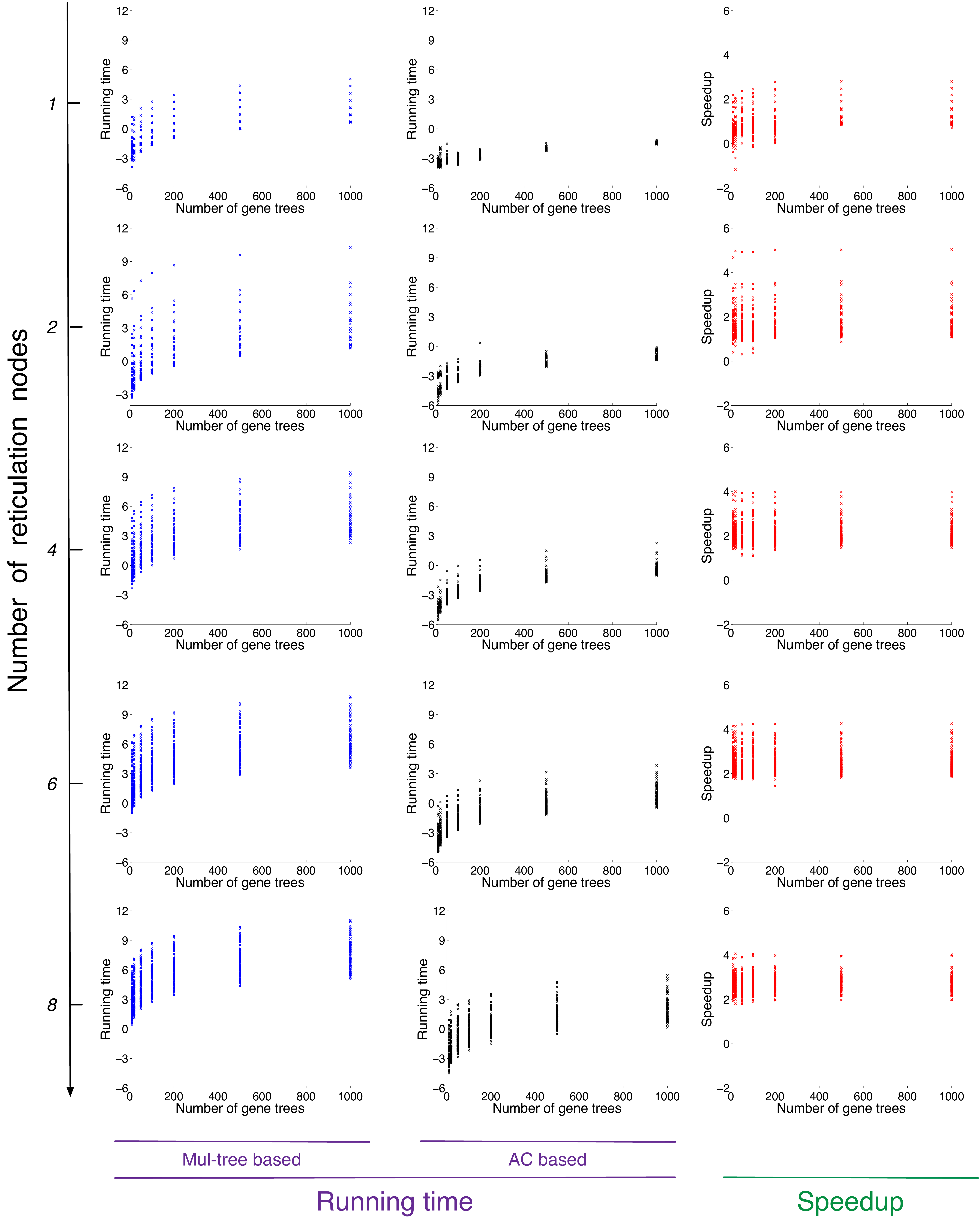}
  \renewcommand{\baselinestretch}{1.0}
   \vspace{-3mm}
   \caption{\small The running times ($\ln$ of number seconds) of the MUL-tree based ($t(MUL)$), and AC-based ($t(AC)$) methods for computing 
   parsimonious reconciliations, as well as the speedup $log_{10}(t(MUL)/t(AC))$. 
  \label{fig:mdcResult} 
   }
\vspace{-10mm}
  \end{center}
\end{figure*}

 For computing the minimum number of extra lineages, the results of the running time of both two methods are shown in Fig. \ref{fig:mdcResult}. Overall, both two methods spent more time on data sets where the species networks contain more reticulation nodes. It is not surprising given the fact that adding more reticulation nodes increases the complexity of the networks in general. We can see that the speedup of the AC-based method over the MUL-tree based method also increased when the number of reticulation nodes in the species networks increased. It is up to over $5$ orders of magnitude. In this figure, we only plot the results of the computations that could finish in $24$ hours across all different number of loci sampled. In fact, the AC based method finished every computation in less than 3 minutes, even for the largest data set which contained species networks with $8$ reticulations and $1000$ gene trees. For the MUL-tree based one, out of $100$ repetitions the numbers of repetitions that were able to finish in $24$ hours across all different loci are $100$, $100$, $99$, $96$ and $88$ for data sets containing species networks with $1$, $2$, $4$, $6$ and $8$ reticulation nodes. 

\begin{figure*}[!ht]   
  \begin{center}
   \vspace{-2mm}
   \includegraphics[width=0.8\textwidth]{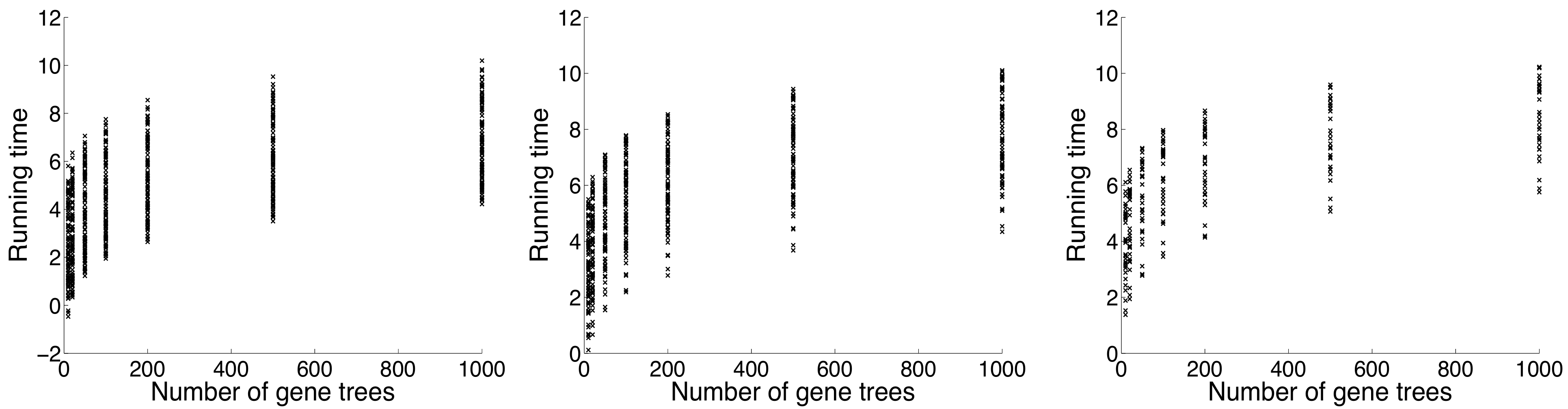}
  \renewcommand{\baselinestretch}{1.0}
   \vspace{-3mm}
   \caption{\small The running time ($\ln$ of number of seconds) of the AC-based algorithm for computing the probability of gene tree topologies given a species network. The columns from left to right correspond to data sets containing species networks with $1$, $4$ and $8$ reticulation nodes, respectively. \label{fig:probResult} 
   }
\vspace{-10mm}
  \end{center}
\end{figure*}

For computing the probability of the gene tree topologies given a species network,  we were not able to run the MUL-tree based one because we found it could not finish the computation in $24$ hours given even for the smallest data set (one gene tree and a species network with one reticulation node). In contrast, the AC-based method only needed $0.4$ seconds on the same data set which implies a speedup of at least $5$ orders of magnitude. Part of the results of the AC based algorithm are shown in Fig. \ref{fig:probResult}. 
 Again, only the results of the computations that could finish successfully in 24 hours across all loci were plotted. We can see that the number of data points in the figure decreased significantly when the number of reticulation nodes in the species networks increased. In fact, out of $100$ repetitions, the numbers of repetitions that finished the computations successfully across all different loci are $99$, $96$, $84$, $54$ and $32$ for data sets containing species networks with $1$, $2$, $4$, $6$ and $8$ reticulation nodes respectively. The number of successful runs is much smaller than that for the parsimony method. Furthermore, those computations failed not only because of the 24 hours time limit. Part of them are due to memory issues:  the number of configurations generated in the computation in order to cover all the possible coalescence patterns that could arise is much more than that needed in the parsimony method. And the increase in the number of reticulation nodes in the species network might result in a very large increase in the number of configurations.

From Fig. \ref{fig:mdcResult} and Fig. \ref{fig:probResult} we observe that for both methods, the running time differed significantly 
from one data set to another.  There are several factors that can affect the number of configurations generated during the computation which directly dominates the running time of the algorithm.  Two of the factors that affect performance are the number of leaves under 
a reticulation node, as well as the topology of the gene tree. We considered a ``controlled" data set, where we controlled the placement of the reticulation node as well as the shapes of the gene trees. In particular, we considered three networks, each 
 with a single reticulation node, yet with 1, 8, and 15 leaves under the reticulation node, respectively (see Fig.~\ref{fig:analysis1} in 
 the Appendix). Further, we considered two gene trees: $gt_1$, whose topology is ``contained" with each of the three networks, and 
 $gt_2$, whose disagreement with the three phylogenetic networks is very extensive that all coalescence events must occur above 
 the root of the phylogenetic networks (Fig.~\ref{fig:analysis1} in the Appendix). We ran both AC-based methods on every pair of 
 phylogenetic network and gene tree. 

 For the parsimony method, if the gene tree is a contained tree of the species network, it can be reconciled into the species network with $0$ extra lineages. In this case, for every articulation node $v$ of the network, $\AC_v$ has only one element $(AC,w)$ and $n(AC)=1$, and the running time is almost the same for all three networks and it is very fast (Table \ref{table:analysis1_MP} in the Appendix).   However, for gene tree $g_2$ whose coalescent events have to happen all above the root, for every articulation node $v$, $\AC_v$ has only one element $(AC,w)$ and $n(AC)=q$ where $q$ equals the number of leaf nodes under $v$. We know that at a reticulation node every configuration $AC$ will give rise to $2^{n(AC)}$ configurations to each of its parents. Therefore, the 
   running time of $g_2$ increased when the number of nodes under the reticulation nodes in the species network increased, and $m$ who is a parent of the reticulation node $h$ has the largest $\AC_v$ set and $|\AC_v| = q$ where $q$ is the number of leaves under $h$ (Table \ref{table:analysis1_MP} in the Appendix). Furthermore, we found that the number of valid allele mappings when using the MUL-tree based method is equal to the largest size of $\AC_v$ generated for a node $v$ during the computation when all the coalescent events have to happen above the root of the species network if we do not reduce the number of configurations for articulation nodes. This is easy to see. For the AC-based algorithm, if we do not clear the splitting information at articulation nodes, then every element in $\AC_R$, where $R$ is the root of the network, represents a different combination of the ways every leaf lineage took at every reticulation node. And every valid allele mapping also represents the same thing. However, for most of the gene trees, not all coalescent events have to happen above the root, and that is part of where the AC-based algorithm improves upon the 
   MUL-tree based one. 
 Comparing $g_1$ and $g_2$ we can see that for parsimony reconciliations, the more coalescent events that are allowed to occur under reticulation nodes with respect to the topology of the gene tree, the faster the method is.

For the probabilistic method, since we need to keep all configurations so as to cover all possible coalescence patterns, the gene trees whose coalescent events have to happen above the root become the easiest case because they have only one reconciliation. It is exactly the opposite to the parsimony method where the gene trees whose coalescent events have to happen above the root take longest running time (Table \ref{table:analysis1_ML} in the Appendix). 
 For the MUL-tree based method, the probability is computed by summing up the probabilities of all coalescent histories in MUL-tree under all valid allele mappings. However, for most cases using ACs to compute the probability of a gene tree given a species tree is much faster than through enumerating coalescent histories due to the fact that the number of coalescent histories is  much larger than the number of configurations generated \cite{Wu12}. That is part of the reason why the AC based algorithm outperforms the MUL-tree based one for computing the probability in terms of efficiency.

A third factor that impacts performance is the dependency of the reticulation nodes in the phylogenetic network (roughly, how many 
of them fall on a single path to the root).  For parsimonious reconciliations, when the reticulation events are independent (or, less
 dependent), the method is much faster. This is not surprising, given that almost all nodes are articulation nodes and the number of ACs is reduced significantly. For the probabilistic reconciliation, a similar trend holds, and the dependence of the reticulation nodes 
 results in an explosion in the number of ACs. These results are given in more detail in Fig.~\ref{fig:analysis2} and Tables \ref{table:analysis2_MP} and \ref{table:analysis2_ML} in the Appendix.

To sum up, for the data sets of the same size (e.g., number of taxa and  reticulation nodes), the running time of the AC-based algorithms increases when there are more leaves under reticulation nodes and when the reticulation nodes are more dependent on each other. With respect to the topology of the gene tree and the species network, the more coalescent events that are allowed under reticulation nodes the faster the parsimony method is, and the opposite for the probabilistic method. For most cases, the AC-based methods are significantly much faster than the MUL-tree based ones. For parsimony, the gain in terms of efficiency comes from avoiding considering useless allele mappings including the ones that cannot yield the optimal 
 reconciliation implied by the coalesced lineages in the configurations and the ones that correspond to the configurations being removed at articulation nodes. For probabilistic reconciliation, the gain comes from two parts. One is also avoiding considering useless allele mappings by removing corresponding configurations at articulation nodes. The other is using AC to compute the probability instead of enumerating the coalescent histories.


\newpage
\section*{APPENDIX}

\begin{figure*}[ht]   
  \begin{center}
   \vspace{-2mm}
   \includegraphics[width=0.4\textwidth]{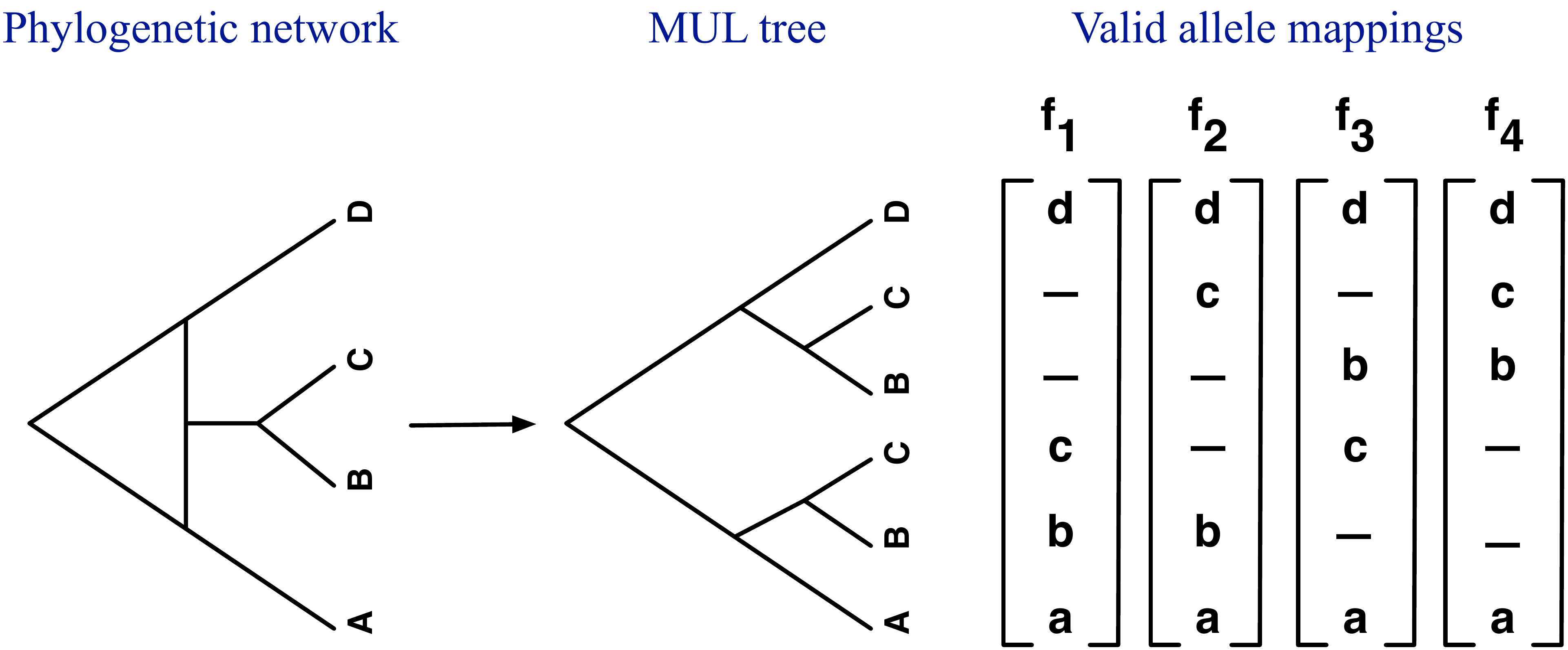}
  \renewcommand{\baselinestretch}{1.0}
   \vspace{-3mm}
   \caption{\small Illustration of the conversion from a phylogenetic network to a MUL-tree, along with all valid allele mappings associated with the case in which single alleles $a$, $b$, $c$ and $d$ were sampled from each of the four species $A$, $B$, $C$ and $D$, respectively. \label{fig:Net2Tree} 
   }
\vspace{-4mm}
  \end{center}
\end{figure*}

\begin{figure*}[!ht]   
  \begin{center}
   \vspace{-2mm}
   \includegraphics[width=0.8\textwidth]{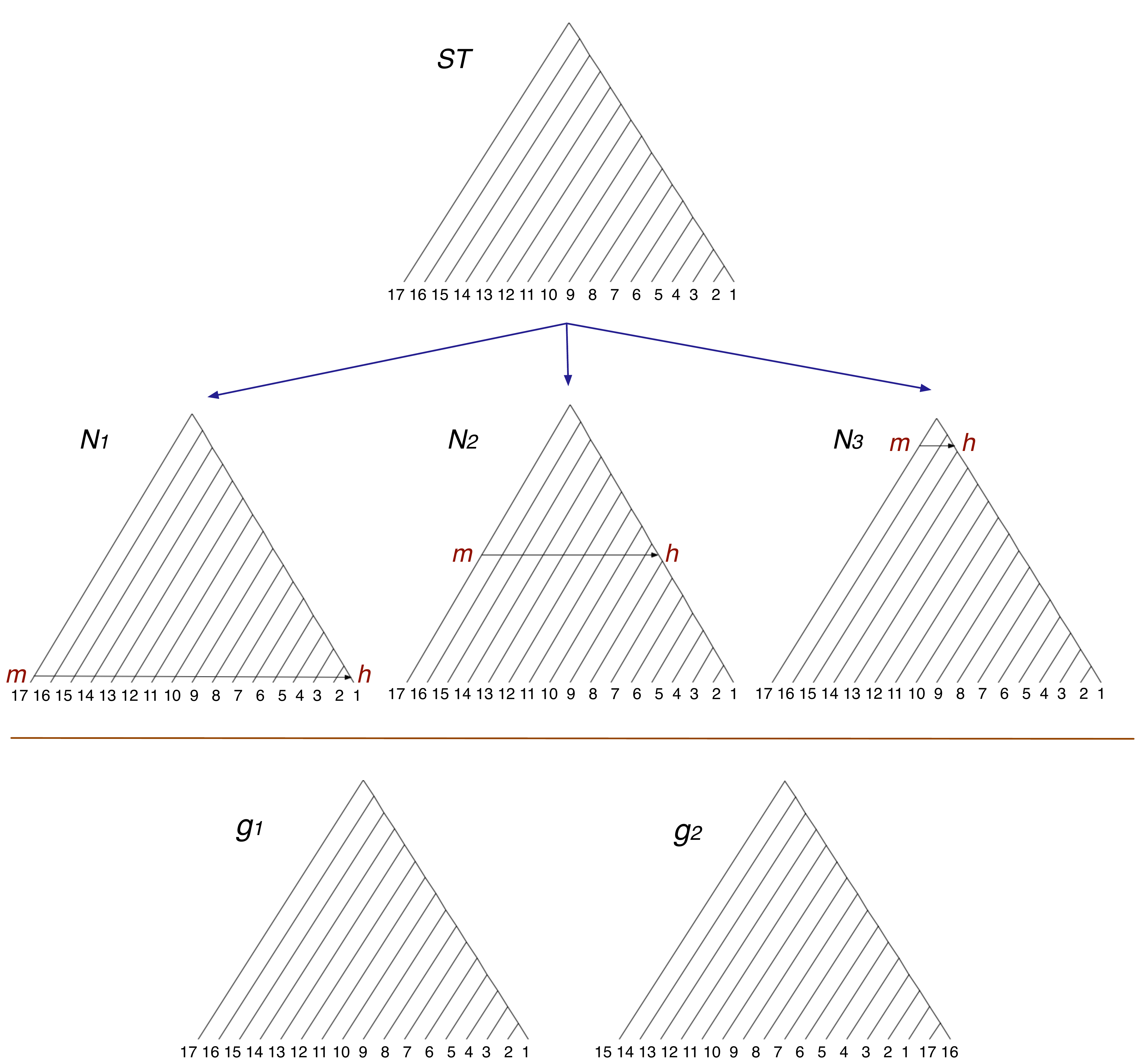}
  \renewcommand{\baselinestretch}{1.0}
   \vspace{-3mm}
   \caption{\small Synthetic data with controlled placements of the reticulation nodes. (Top) A species tree $ST$. 
   (Middle) $N_1$, $N_2$ and $N_3$ are three species networks constructed by adding one reticulation edge to $ST$ at three different locations. (Bottom) Two gene trees $g_1$, which is contained in all three networks, and $g_2$, whose coalescent events have to happen above the root of all three networks. \label{fig:analysis1} 
   }
\vspace{-4mm}
  \end{center}
\end{figure*}

\begin{table}[!ht]
\footnotesize
\caption{\footnotesize The results of running the AC-based algorithm for computing the minimum number of extra lineages given gene trees and species networks in Fig. \ref{fig:analysis1}. $|\AC_h|$ is the number of configurations at the reticulation node $h$ and $max{|\AC|}$ is the maximum number of configurations generated at a node during computation. We labeled the first node $v$ in post-order of traversal that contains the largest $AC_v$ set by $m$ in Fig. \ref{fig:analysis1}. Furthermore, the last column is the number of valid allele mappings if using the MUL-tree based method.\label{table:analysis1_MP}} 
\centering
\begin{tabular}{|c|c|c|c||c|c|c||c|}
\hline
 &  \multicolumn{3}{c||}{$g_1$}  &  \multicolumn{3}{c||}{$g_2$} &\\
  \cline{2-7}
 & $|\AC_h|$ & $max{|\AC|}$ & running time (s) & $|\AC_h|$ & $max{|\AC|}$ & running time (s) & $\#$allele mappings\\
 \hline
$N_1$ & $1$ & $2$ & $0.011$ & $1$ & $2$ & $0.016$ & $2$ \\
$N_2$ & $1$ & $2$ & $0.013$ & $1$ & $256$ $(2^8)$ & $0.105$ & $256$ \\
$N_3$ & $1$ & $2$ & $0.013$ & $1$ & $32768$ $(2^{15})$ & $32.551$ & $32768$ \\
\hline
\end{tabular}
\end{table}

\begin{table}[!ht]
\footnotesize
\centering
\caption{\footnotesize The results of running the AC-based algorithm for computing the probability of gene tree topologies given gene trees and species networks in Fig. \ref{fig:analysis1}. $|\AC_h|$ is the number of configurations at the reticulation node $h$ and $max{|\AC|}$ is the maximum number of configurations generated at a node during computation. We labeled the first node $v$ in post-order of traversal that contains the largest $AC_v$ set by $m$ in Fig. \ref{fig:analysis1}. Furthermore, the last column is the number of valid allele mappings if using the MUL-tree based method. \label{table:analysis1_ML}}
\begin{tabular}{|c|c|c|c||c|c|c||c|}
\hline
 &  \multicolumn{3}{c||}{$g_1$}  &  \multicolumn{3}{c||}{$g_2$} &\\
  \cline{2-7}
 & $|\AC_h|$ & $max{|\AC|}$ & running time (s) & $|\AC_h|$ & $max{|\AC|}$ & running time (s) & $\#$allele mappings\\
 \hline
$N_1$ & $1$ & $16$ & $0.075$ & $1$ & $2$ & $0.019$ & $2$ \\
$N_2$ & $8$ & $813$ & $0.526$ & $1$ & $256$ $(2^8)$ & $0.232$ & $256$ \\
$N_3$ & $15$ & $98286$ & $617.845$ & $1$ &  $32768$ $(2^{15})$ & $34.968$ & $32768$ \\
\hline
\end{tabular}
\end{table} 

\begin{figure*}[!ht]   
  \begin{center}
   \vspace{-2mm}
   \includegraphics[width=0.8\textwidth]{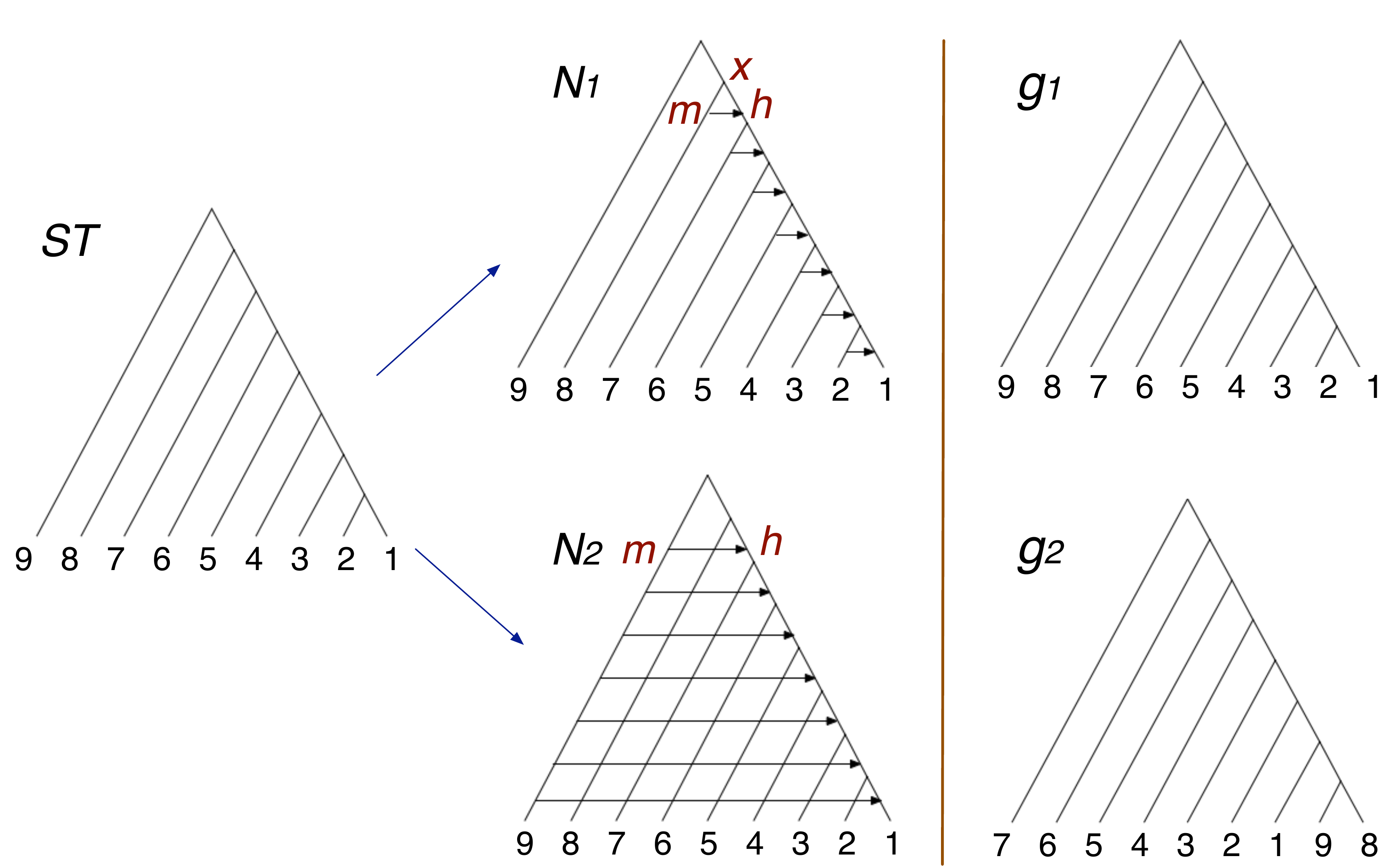}
  \renewcommand{\baselinestretch}{1.0}
   \vspace{-3mm}
   \caption{\small The effects of dependency of reticulation nodes in the species network and different gene tree topologies on the running time of the AC-based algorithms. (Left) A species tree $ST$. (Middle) $N_1$ and $N_2$ are two species networks constructed by adding seven reticulation edges to $ST$ at different locations. (Right) two gene trees $g_1$, which is a contained tree of both $N_1$ and $N_2$, and $g_2$ whose coalescent events have to happen above the root of both two species networks. \label{fig:analysis2} 
   }
\vspace{-4mm}
  \end{center}
\end{figure*}

\begin{table}[!ht]
\footnotesize
\centering
\caption{\footnotesize The results of running the AC-based algorithm for computing the minimum number of extra lineages given gene trees and species networks in Fig. \ref{fig:analysis2}. $|\AC_h|$ is the number of configurations at the highest reticulation node $h$ and $max{|\AC|}$ is the maximum number of configurations generated at a node during computation. We labeled the first node $v$ in post-order of traversal that contains the largest $AC_v$ set by $m$ in Fig. \ref{fig:analysis2}. Furthermore, the last column is the number of valid allele mappings if using the MUL-tree based method. \label{table:analysis2_MP}} 
\begin{tabular}{|c|c|c|c||c|c|c||c|}
\hline
 &  \multicolumn{3}{c||}{$g_1$}  &  \multicolumn{3}{c||}{$g_2$} &\\
  \cline{2-7}
 & $|\AC_h|$ & $max{|\AC|}$ & running time (s) & $|\AC_h|$ & $max{|\AC|}$ & running time (s) & $\#$allele mappings\\
 \hline
$N_1$ & $1$ & $2$ & $0.014$ & $1$ & $128$ $(2^7)$ & $0.039$ & $268435456$ \\
$N_2$ & $874$ & $5914$ & $2.85$ & $5040$ & $40320$ & $55.684$ & $40320$ \\
\hline
\end{tabular}
\end{table}

\begin{table}[ht]
\footnotesize
\centering
\caption{\footnotesize The results of running the AC-based algorithm for computing the probability of gene tree topologies given gene trees and species networks in Fig. \ref{fig:analysis2}. $|\AC_h|$ is the number of configurations at the highest reticulation node $h$ and $max{|\AC|}$ is the maximum number of configurations generated at a node during computation. We labeled the first node $v$ in post-order of traversal that contains the largest $AC_v$ set by $m$ in Fig. \ref{fig:analysis2}. Furthermore, the last column is the number of valid allele mappings if using the MUL-tree based method. \label{table:analysis2_ML}}
\begin{tabular}{|c|c|c|c||c|c|c||c|}
\hline
 &  \multicolumn{3}{c||}{$g_1$}  &  \multicolumn{3}{c||}{$g_2$} &\\
  \cline{2-7}
 & $|\AC_h|$ & $max{|\AC|}$ & running time (s) & $|\AC_h|$ & $max{|\AC|}$ & running time (s) & $\#$allele mappings\\
 \hline
$N_1$ & $7$ & $274$ & $0.26$ & $1$ & $128$ $(2^7)$ & $0.124$ & $268435456$ \\
$N_2$ & $9928$ & $146433$ & $1336.494$ & $5040$ & $40320$ & $57.418$ & $40320$ \\
\hline
\end{tabular}
\end{table}

\end{document}